\documentclass[11pt]{article}
\newcommand{\ket}[1]{\left | #1 \right \rangle}

\newcommand{\mod}{\mbox{mod }}
\newcommand{\beq}{\begin{equation}}
\newcommand{\eeq}{\end{equation}}
\newcommand{\beqa}{\begin{eqnarray}}
\newcommand{\eeqa}{\end{eqnarray}}

\def\cz{{\cal Z}}

\begin{document}
\begin{center}
{\Large\bf Quantum factoring, discrete logarithms \\ and the hidden
subgroup problem }\\
\bigskip
{\normalsize Richard Jozsa }\\
\bigskip
{\small\it Department of Computer Science, University of Bristol,\\
Woodland Road, Bristol BS8 1UB U.K. }
\\[4mm]

\end{center}

\begin{abstract}
Amongst the most remarkable successes of quantum computation are
Shor's efficient quantum algorithms for the computational tasks of
integer factorisation and the evaluation of discrete logarithms. In
this article we review the essential ingredients of these algorithms
and draw out the unifying generalization of the so-called abelian
hidden subgroup problem. This involves an unexpectedly harmonious
alignment of the formalism of quantum physics with the elegant
mathematical theory of group representations and fourier transforms
on finite groups. Finally we consider the non-abelian hidden subgroup
problem mentioning some open questions where future quantum
algorithms may be expected to have a substantial impact.

\end{abstract}

\section{Introduction}

Quantum algorithms exploit quantum physical effects to provide new
modes of computation which are not available to ``conventional''
(classical) computers. In some cases these modes provide efficient
(i.e. polynomial time) algorithms for computational tasks where no
efficient classical algorithm is known. The most celebrated
quantum algorithm to date is Shor's algorithm for integer
factorisation \cite{SHO94,EKE96,JOZ98b}. It provides a method for
factoring any integer of $n$ digits in time (i.e. in a number of
computational steps) that grows less rapidly than $O(n^3 )$. Thus
it is a polynomial time algorithm in contrast to the best known
classical algorithm for this fundamental problem, which runs in
superpolynomial time of order exp$(n^{\frac{1}{3}} (\log
n)^{\frac{2}{3}})$.

At the heart of the quantum factoring algorithm is the discrete
Fourier transform and the remarkable ability of a quantum computer to
efficiently determine periodicities. This in turn rests on the
mathematical formalism of fast Fourier transforms combined with
principles of quantum physics. In this article we will review these
issues including further applications such as the evaluation of
discrete logarithms. We will outline a unifying generalization of
these ideas: the so-called hidden subgroup problem which is just a
natural group theoretic generalization of the problem of periodicity
determination. Finally we will consider some interesting open
questions related to the hidden subgroup problem for non-abelian
groups, where future quantum algorithms may be expected to have a
substantial impact.

We may think of periodicity determination as a particular kind of
pattern recognition. Quantum computers are able to store and process
large volumes of information, represented compactly in the identity
of an entangled quantum state, but quantum measurement theory
severely restricts our access to the information. Indeed only a
relatively small amount of the information may be read out but this
may be of a ``global'' nature, such as a few broad features of a
large intricate pattern, which may be impossible to extract
efficiently by classical means. This intuition is exemplified in the
earliest quantum algorithm, known as Deutsch's algorithm
\cite{JOZ98b}. Here we are given a black box that computes a Boolean
function of $n$ variables (i.e. a function of all $n$ bit strings
with one-bit values). It is promised that the function is either a
constant function or `balanced' in the sense that exactly half of the
values are 0 and half are 1. We wish to determine with certainty
whether the given function is balanced or constant, using the least
number of queries to the box. Thus we are asking for one bit of
information about the $2^n$ values of the function. Classically
$2^{n-1}+1$ queries are necessary in the worst case (if the problem
is to be solved with {\em certainty}) but quantumly the problem can
be solved in all cases with just one query \cite{JOZ98b}. However if
we tolerate any arbitrarily small probability of error in the answer
then there is also a classical algorithm using only a constant number
of queries.

Inspired by these results, Simon \cite{SIM94} considered a more
complicated situation of a class of functions from $n$ bits to $n$
bits and developed a computational task displaying an exponential gap
between the classical and quantum query complexities, even if (in
contrast to Deutsch's algorithm) the algorithm is required to work
only with bounded error probability of $1/3$ i.e. we allow
probabilistic algorithms and in any run the answer must be correct
with probability at least $2/3$.

In retrospect (c.f. below) Simon's problem turns out to be an
example of a ``generalized periodicity'' or hidden subgroup
problem, for the group of $n$ bit strings under binary bitwise
addition. Shor recognized the connection with periodicity
determination and generalized the constructions to the group of
integers modulo $N$, showing significantly that the associated
discrete Fourier transform may be efficiently implemented in that
context as well. Finally using known reductions of the tasks of
integer factorisation and evaluation of discrete logarithms to
periodicity determinations, he was able to give polynomial time
quantum algorithms for these computational tasks too.

\section{The quantum Fourier transform and periodicities}

We begin with an account of how a quantum computer may efficiently
determine the periodicity of a given periodic function. Consider the
following basic example. Suppose that we have a black box which
computes a function $f:{\cal Z}_N \rightarrow {\cal Z}$ that is
guaranteed to be periodic with some period $r$:
\begin{equation} \label{C3S2R} f(x+r)=f(x) \hspace{1cm} \mbox{
for all $x$} \end{equation} Here ${\cal Z}_N$ denotes the additive
group of integers modulo $N$. We also assume that $f$ does not take
the same value twice within any single period. Note that eq.
(\ref{C3S2R}) can hold only if $r$ divides $N$ exactly.

Our aim is to determine $r$. Classically (in the absence of any
further information about $f$) we can merely try different values
of $x$ in the black box hoping for two equal results which will
then give information about $r$. Generally we will require $O(N)$
random tries to hit two equal values with high probability. Using
quantum effects we will be able to find $r$ using only $O((\log
N)^2)$ steps, which represents an exponential speedup over any
known classical algorithm.

In the quantum context we assume the black box is a coherent quantum
process which evolves the input state $\ket{x}\ket{0}$ to
$\ket{x}\ket{f(x)}$ i.e. the values of $x$ and $f(x)$ are labels on a
suitable set of orthogonal states. We begin by computing all values
of $f$ in equal superposition, using one application of the box. To
do this we set up the input register in the equal superposition
$\frac{1}{\sqrt{N}} \sum_x \ket{x}$, apply the function and obtain
the state:
\begin{equation} \label{C3S2FX} \ket{f} = \frac{1}{\sqrt{N}}
\sum_{x=0}^{N-1} \ket{x}\ket{f(x)} \end{equation} Although the
description of this state embodies all the values of $f$ and hence
the periodicity, it is not immediately clear how to extract the
information of $r$! If we measure the value in the second
register, giving a value $y_0$ say, then the state of the first
register will be reduced to an equal superposition of all those
$\ket{x}$'s such that $f(x)=y_0$. If $x_0$ is the least such $x$
and $N=Kr$ then we will obtain in the first register the periodic
state
\begin{equation} \label{C3S2REG1}
\ket{\psi}= \frac{1}{\sqrt{K}} \sum_{k=0}^{K-1} \ket{x_0 + kr}
\end{equation}
It is important to note here that $0\leq x_0 \leq r-1$ has been
generated at random, corresponding to having seen any value $y_0$
of $f$ with equal probability. So if we now measure the value in
this register, the overall result is merely to produce a number
between 0 and $N-1$ uniformly at random, giving no information at
all about the value of $r$.

The resolution of this difficulty is to use the Fourier transform
which, even for classical data, is known to be able to pick out
periodic patterns in a set of data regardless of how the whole
pattern is shifted. The discrete Fourier transform $\cal F$ for
integers modulo $N$ is the $N$ by $N$ unitary matrix with entries
\begin{equation} \label{C3S2FT} {\cal F}_{ab}  =
\frac{1}{\sqrt{N}} e^{2\pi i\frac{ab}{N}} = \frac{1}{\sqrt{N}}
\chi_a (b)
\end{equation}
where we have introduced the functions \begin{equation}
\label{chis} \chi_l (m) = \exp 2\pi i \frac{lm}{N}. \end{equation}
If we apply this unitary transform to the state $\ket{\psi}$ above
then we obtain \cite{EKE96}
\begin{equation} {\cal F} \ket{\psi} = \frac{1}{\sqrt{r}}
\sum_{j=0}^{r-1} e^{2\pi i \frac{x_0 j}{r} } \ket{j\frac{N}{r}}
\end{equation}
Indeed a direct calculation shows that the labels which appear
with non-zero amplitude are those values of $l$ satisfying
\begin{equation} \label{chione} \chi_l (r) = e^{2\pi i
\frac{lr}{N}} = 1 \end{equation} i.e. $lr$ is a multiple of $N$ and
furthermore they appear with equal squared amplitudes. This
calculation uses the periodic structure of eq. (\ref{C3S2REG1}) and
the elementary identity
\begin{equation} \label{elem} \sum_{k=0}^{K-1} \left( e^{2\pi i \frac{l}{K}}
\right)^k = \left\{ \begin{array}{l} \mbox{$0$ if $l$ is not a
multiple of $K$} \\ \mbox{$K$ if $l$ is a multiple of $K$}
\end{array} \right. \end{equation}

It is important to note here that the random shift $x_0$ no longer
appears in the ket labels. If we now read the label we will obtain a
value $c$ say, which is necessarily a multiple of $N/r$ i.e.
$c=\lambda N/r$ . Thus we can write
\begin{equation}
\frac{c}{N} = \frac{\lambda}{r} \end{equation} where $c$ and $N$ are
known numbers and and $0\leq \lambda \leq r-1$ has been chosen
uniformly at random by the measurement. Now if the randomly chosen
$\lambda$ is fortuitously coprime to $r$ (i.e. $\lambda$ and $r$ have
no common factors) we can determine $r$ by cancelling $c/N$ down to
an irreducible fraction. What is the probability that a randomly
chosen $r$ actually {\em is} coprime to $r$? According to a basic
theorem of number theory (c.f. \cite{HAR65, SCH90} and appendix A of
\cite{EKE96}), the number of co-primes less than $r$ goes as
$e^{-\gamma}r/\log \log r$ (where $\gamma$ is Euler's constant) for
large $r$. Thus the probability that our randomly chosen $\lambda$ is
coprime to $r$ is $O(1/\log\log r)$ which exceeds $O(1/\log\log N)$.
Hence if we repeat the above procedure $O(\log\log N)$ times we can
succeed in determining $r$ with any prescribed probability
$1-\epsilon$ as close to 1 as desired.

We noted above that we want our quantum algorithm to run in time
poly($\log N$) i.e. in a number of steps which is polynomial in $\log
N$ rather than $N$ itself, to achieve an exponential speed up over
any known classical algorithm for determining periodicity. We showed
above that merely $O(\log\log N)$ repetitions suffice to determine
$r$ but there is still a significant gap in our argument: the Fourier
transform $\cal F$ that we used is a large non-trivial unitary
operation, of size $N$ by $N$, and we cannot {\em ab initio} just
assume that it can be implemented using only poly ($\log N$) basic
computational operations. Indeed it may be shown that any $d$ by $d$
unitary operation may be implemented on a quantum computer (equipped
with any universal set of operations) in $O(d^2 )$ steps
\cite{EKE96}. This is also the number of steps needed for the
classical computation of multiplying a $d$ by $d$ matrix into a $d$
dimensional column vector. For our use of $\cal F$ this bound of
$O(N^2 )$ does not suffice. Fortunately the Fourier transform (FT)
has extra special properties which enable it to be implemented in
$O((\log N)^2 )$ steps. These properties stem from the classical
theory of the fast Fourier transform (FFT) \cite{MAS95} which shows
how to reduce the $O(N^2 )$ steps of classical matrix multiplication
to $O(N\log N )$ steps. If the same ideas are implemented in a
quantum setting then it may be seen \cite{EKE96,EKE98} that the
number of steps is reduced to $O((\log N )^2 )$ giving our desired
implementation. Note also that according to eq. (\ref{C3S2FT}) we
have
\[ {\cal F} \ket{0} = \frac{1}{\sqrt{N}} \sum_{x=0}^{N-1} \ket{x} \]
so that once we have an efficient implementation of $\cal F$ we will
be able to efficiently produce the uniform large superposition in the
input register, necessary to get $\ket{f}$ in eq. (\ref{C3S2FX}).

The technical details of the efficient implementation of FT are given
in \S 3 of \cite{EKE98} but the essential idea is the following. We
will be able to efficiently implement FT in dimensions which are
powers of 2 rather than arbitrary $N$. Thus we use the smallest power
of 2 that is larger than $N$. (In later applications this slight
mismatch of dimensions can be shown to not cause problems, although
the rigorous demonstration of this \cite{SHO94} can become
technically complicated). Let $n$ denote the least integer greater
than $\log_2 N$. Then the required Fourier transform FT is a unitary
operation on $n$ qubits. The FFT formalism gives an explicit way of
decomposing FT on $n$ qubits into a sequence of gates where each gate
acts on at most two qubits and the length of the sequence is
polynomial in $n$ (actually $O(n^2 )$). FT is a very special
operation in this regard -- a general unitary operation would require
a sequence of {\em exponential} length! Consider now the action of a
2-qubit gate $U$ on a state $\ket{\alpha}$ of $n$ qubits. Suppose
that $U$ acts on the first two qubits and that $U$ has matrix
elements $U^{i_1 i_2}_{j_1 j_2}$ in a standard product basis of the
$n$ qubit state space. Suppose that $\ket{\alpha}$ has components
$a_{i_1 \ldots i_n }$ (where all indices range over the values 0 and
1). The components of the updated state are given by matrix
multiplication: \beq \label{new} a_{i_1 \ldots i_n }^{new} =
\sum_{j_1 , j_2} U^{j_1 j_2}_{i_1 i_2} a_{j_1 j_2  i_3 \ldots i_n }.
\eeq This update counts as one step of quantum computation (or more
precisely a constant number, independent of $n$ to implement $U$) and
the FFT decomposition amounts to an implementation of FT in $O(n^2 )$
steps on a quantum computer. In contrast if eq. (\ref{new}) is viewed
as a classical computation, we must perform a $4\times 4$ matrix
multiplication $2^{n-2}$ times (for all values of the string $i_3
\ldots i_n$). This ultimately gives an implementation of FT with
$O(n2^n )$ classical steps, which is the standard fast Fourier
transform algorithm.

In summary, the quantum algorithm for determining the periodicity of
a given function $f$,  with $N$ inputs, begins with the computation
of all values of $f$ in superposition using one application of FT and
one evaluation of $f$. FT is then applied to pick out the periodic
structure of the resulting state. The quantum implementation of the
FFT algorithm guarantees that FT may be implemented in poly($\log N$)
steps. An analogous classical computation would require $O(N)$
invocations of $f$ to compute a column vector of all the function
values and then $O(N\log N)$ steps to perform the FFT. Thus the
quantum algorithm represents an exponential speedup.

\section{Quantum factoring}

The problem of integer factorisation is the following: given a number
$N$, of $n=\log_2 N$ digits, we wish to determine a number $k$ (not
equal to 1 or $N$) which divides $N$ exactly. We now outline how this
problem may be reduced to a problem of periodicity determination for
a suitable periodic function $f$. Then the quantum algorithm
described in the preceding section will achieve the factorisation of
$N$ in poly($n$) time i.e. polynomial in the number of digits of $N$.

We note first that there is no known classical algorithm which will
factorise any given $N$ in a time polynomial in the number of digits
of $N$. For example the most naive factoring algorithm involves
test-dividing $N$ by each number from 1 to $\sqrt{N}$ (as any
composite $N$ must have a factor in this range). This requires at
least $\sqrt{N}$ steps (at least one step for each trial factor) and
$\sqrt{N}= 2^{\frac{1}{2}n}$ is exponential in $n$. In fact using all
the ingenuity of modern mathematics, the fastest known classical
factoring algorithm runs in a time of order exp$(n^{\frac{1}{3}}
(\log n)^{\frac{2}{3}})$.

To reduce the problem to a problem of periodicity we will need to use
some basic results from number theory. These are further described in
the appendix of \cite{EKE96} and complete expositions may be found in
most standard texts on number theory such as \cite{HAR65,SCH90}. We
begin by selecting a number $a<N$ at random. Using Euclid's
algorithm, we compute in poly($\log N$) time, the highest common
factor of $a$ and $N$. If this is larger than 1, we will have found a
factor of $N$ and we are finished! However it is overwhelmingly
likely that a randomly chosen $a$ will be coprime to $N$ (e.g. if $N$
is the product of two large primes). If $a$ is coprime to $N$, then
Euler's theorem of number theory guarantees that there is a power of
$a$ which has remainder 1 when divided by $N$. Let $r$ be the
smallest such power:
\begin{equation} \label{C3S2XR} a^r \equiv 1 \,\,\, \mbox{mod $N$
\hspace{4mm} and $r$ is the least such power} \end{equation} (If
$a$ is not coprime to $N$ then no power of $a$ has remainder 1).
$r$ is called the {\em order} of $a$ modulo $N$. Next we show that
the information of $r$ can provide a factor of $N$.

Suppose that we have a method for determining $r$ (c.f. later) and
suppose further that $r$ comes out to be an {\em even} number. Then
we can rewrite eq. (\ref{C3S2XR}) as $a^r -1 \equiv 0 \,\,\,
\mbox{mod $N$} $ and factorise as a difference of squares:
\begin{equation} \label{C3S2X}
(a^{r/2}-1)(a^{r/2}+1) \equiv 0 \,\,\, \mbox{mod $N$}
\end{equation}
Let $\alpha = a^{r/2}-1$ and $\beta = a^{r/2} +1$. Then $N$ exactly
divides the product $\alpha \beta$. If neither $\alpha$ nor $\beta$
is a multiple of $N$ then $N$ must divide partly into $\alpha$ and
partly into $\beta$. Thus computing the highest common factor of $N$
with $\alpha$ and $\beta$ (again using Euclid's algorithm) will
generate a non-trivial factor of $N$.

As an example take $N=15$ and choose the coprime number $a=7$. By
computing the powers of 7 modulo 15 we find that $7^4 \equiv 1 \,\,\,
\mbox{mod 15}$ i.e. the order of 7 modulo 15 is 4. Thus 15 must
exactly divide the product $(7^{4/2}-1)(7^{4/2}+1) = (48)(50)$.
Computing the highest common factor of 15 with 50 and 48 gives 5 and
3 respectively, which are indeed nontrivial factors of 15.

Our method will give a factor of $N$ provided that $r$ comes out to
be even and that neither of $(a^{r/2}\pm 1)$ are exact multiples of
$N$. To guarantee that these conditions occur often enough (for
randomly chosen $a$'s) we have\\Theorem: Let $N$ be odd and suppose
that $a<N$ coprime to $N$ is chosen at random. Let $r$ be the order
of $a$ modulo $N$. Then the probability that $r$ is even and
$a^{r/2}\pm 1$ are not exact multiples of $N$ is always $\geq
\frac{1}{2}$.\\ The (somewhat lengthy) proof of this theorem may be
found in appendix B of \cite{EKE96}, to which we refer the reader for
details.

Overall, our method will produce a factor of $N$ with probability at
least half in every case. This success probability may be amplified
as close as desired to 1, since $K$ repetitions of the procedure
(with $K$ constant independent of $N$) will succeed in factorising
$N$ with probability exceeding $1-\frac{1}{2^K}$.

All steps in the procedure, such as applying Euclid's algorithm and
the arithmetic manipulation of numbers, can be done in poly($n$)
time. The only remaining outstanding ingredient is a method for
determining $r$ in poly($\log N$) time. Consider the exponential
function:
\begin{equation} \label{C3S2MODEXP} f(x)=a^x\,\,\, \mbox{mod  $N$}
\end{equation}
Now eq. (\ref{C3S2XR}) says precisely that $f$ is periodic with
period $r$ i.e. that $f(x+r)=f(x)$. Thus we use the quantum algorithm
for periodicity determination, described in the previous section, to
find $r$. To apply the algorithm as stated, we need to restrict the
scope of $x$ values in eq. (\ref{C3S2MODEXP}) to a {\em finite} range
$0\leq x\leq q$ for some $q$. If $q$ is not an exact multiple of (the
unknown) $r$ i.e. $q=Ar+t$ for some $0<t <r$, then the resulting
function will not be exactly periodic -- the single final period over
the last $t$ values will be incomplete. However if $q$ is chosen
large enough, giving sufficiently many intact periods of $f$, then
the single corrupted period will have negligible effect on the use of
the $q$ by $q$ Fourier transform to determine $r$, as we might
intuitively expect. In fact it may be shown that if $q$ is chosen to
have size $O(N^2 )$ then we get a reliable efficient determination of
$r$. For the technical analysis of this imperfect periodicity
(involving the theory of continued fractions) we refer the reader to
\cite{SHO94,EKE96}. $q$ is also generally chosen to be a power of 2
to allow an efficient implementation of FT via the FFT formalism.

\section{Evaluation of discrete logarithms}

In the previous section we showed how the problem of factoring may be
reduced to a question of periodicity of a function on ${\cal Z}_N$,
the additive group of integers modulo $N$. We now introduce the
problem of discrete logarithms and show how it may also be reduced to
a slightly more general kind of periodicity -- on the additive group
of pairs of integers modulo $N$. These important special cases
provide the basis for the generalization in the next section to an
elegant and natural group theoretic setting.

Let $p$ be a prime number and let ${\cal Z}_p^*$ denote the group of
integers $\{ 1,2, \ldots ,p-1 \}$ under {\em multiplication} modulo
$p$. Note that for general values of $m$ the set ${\cal Z}_m^* = \{
1,2, \ldots ,m-1 \}$ is not a group under multiplication modulo $m$
as we do not generally have multiplicative inverses (e.g. in ${\cal
Z}_6$ there is no number $x$ satisfying $3x \equiv 1 \,\,\mod 6$ i.e.
3 has no inverse) but if $p$ is prime then ${\cal Z}_p^*$ is always a
group.

A number $g$ in ${\cal Z}_p^*$ is called a generator (or primitive
root $\mod p$) if the powers of $g$ generate all of ${\cal Z}_p^*$
i.e. ${\cal Z}_p^* = \{ g^0 = 1, g^1 ,g^2 , \ldots , g^{p-2} \}$.
(For example in ${\cal Z}_5^*$ 2 and 3 are generators but 1 and 4 are
not). Thus every element $x$ of ${\cal Z}_p^*$ may be written
uniquely as $x= g^y$ for some $y$ in ${\cal Z}_{p-1}$. $y$ is called
the discrete logarithm of $x$ (with respect to $g$) and we write
$y=\log_g x$. Note that multiplication of $x$'s $\mod p$ corresponds
to addition of $y$'s $\mod (p-1)$ so a generator provides a way of
identifying ${\cal Z}_p^*$ as ${\cal Z}_{p-1}$.

The problem of discrete logarithms is the following: we have $p$
and a generator $g$ of ${\cal Z}_p^*$. For any $x \in {\cal
Z}_p^*$ we want to compute its discrete logarithm $y=\log_g x$.
Let $n$ be the number of digits of $p$. The fastest known
classical algorithm runs in time of order exp$(n^{\frac{1}{3}}
(\log n)^{\frac{2}{3}})$ whereas our quantum algorithm will run in
time less than $O(n^3)$.

We begin by noting that multiplicative inverses in ${\cal Z}_p^*$
may be computed efficiently using Euclid's algorithm. Indeed for
any $x$ we have the highest common factor of $x$ and $p$ being 1
so Euclid's algorithm provides integers $a$ and $b$ such that
$ax+bp = 1$ so $ax \equiv 1\,\,\mod p$ and $a$ is the desired
inverse.

Consider $G= \cz_{p-1} \times \cz_{p-1}$, the additive group of pairs
of integers and for given $x,g,p$, the function $f: \cz_{p-1} \times
\cz_{p-1} \rightarrow \cz_p^*$ given by \[ f(a,b) = g^a x^{-b} \,\,
\mod p \] which is computable in time poly$(n)$. In terms of the
discrete logarithm $y=\log_g x$ we have \[ f(a,b) = g^{a-yb} \,\,
\mod p \] so
\[ f(a_1 , b_1 ) = f(a_2,b_2) \mbox{\hspace{2mm} if and only if
\hspace{2mm} $(a_2,b_2) = (a_1,b_1) + \lambda (y,1)$ for $\lambda \in
\cz_{p-1}$}. \] Thus the pair $(y,1)$ is the period of $f$ on its
product domain. To determine $y$ our quantum algorithm will follow
the standard period--finding procedure of section 2, slightly
generalized to deal with the fact that the domain consists of pairs
rather than just single numbers.

We consider a Hilbert space with an orthonormal basis $\{
\ket{a}\ket{b}: a,b \in \cz_{p-1} \}$ labeled by the elements of
$G$ and begin by computing an equal superposition of all values of
$f$: \[ \ket{f} = \frac{1}{p-1} \sum_{a,b}
\ket{a}\ket{b}\ket{f(a,b)}. \] If we measure the last register and
see a value $k_0 = f(a_0,b_0)$ we obtain the periodic state
\[ \ket{\psi} = \frac{1}{\sqrt{p-1}} \sum_{k=0}^{p-2} \ket{a_0
+ky}\ket{b_0 + k}. \] To eliminate the dependence of the labels on
the randomly chosen $(a_0,b_0)$ we apply $\cal F$, the Fourier
transform modulo $(p-1)$ to each of the two registers. The
calculations are very similar to those for factoring (c.f. eq.
(\ref{elem})). Let us introduce the functions
\[ \chi_{l_1,l_2} (a,b)= \exp 2\pi i(\frac{al_1 + bl_2 }{p-1}). \]
Then (similar to eq. (\ref{chione})) ${\cal F}\otimes {\cal F}
\ket{\psi}$ will yield an equally weighted superposition of those
labels $(l_1,l_2)$ such that $\chi_{l_1,l_2}(y,1)=1$ i.e.
$yl_1+l_2\equiv 0 \,\,\mod p-1$ so $l_2 = -yl_1 \,\, \mod p-1$ and
$l_1 = 0,1, \ldots ,p-2$. Explicitly we have
\[ {\cal F}\otimes {\cal F} \ket{\psi} = \frac{1}{\sqrt{p-1}}
\sum_{l_1 =0}^{p-2} \exp 2\pi i (\frac{a_0 l_1 - b_0 yl_1}{p-1})\,\,
\ket{l_1}\ket{-yl_1}. \] Then a measurement of the labels will
provide a pair $(l_1,l_2)=(l_1,-yl_1 \,\, \mod p-1 )$ where $l_1 \in
\cz_{p-1}$ is chosen uniformly at random. If $l_1$ happens to be
coprime to $p-1$ we can use Euclid's algorithm to find $l_1^{-1}$,
the multiplicative inverse modulo $p-1$, and compute $y$ as
$-l_1^{-1}l_2$. If $l_1$ is not coprime to $p-1$ then we cannot
uniquely determine $y$ from $(l_1,l_2)$. What is the probability that
a uniformly chosen $l_1$ is coprime to $p-1$? In section 2 we saw
that this probability will be of order $1/\log \log (p-1)$ and so to
determine $y$ with high probability we will need to repeat our
algorithm a very modest $O(\log \log p)$ times (which is even
exponentially smaller than our goal of poly$(\log p)$ times).

As in the case of factoring there is the residual issue of
efficiently implementing the Fourier transform that is used. To
take advantage of the FFT formalism we would want to use FT for
integers modulo a power of 2 (instead of modulo $p-1$). Let $2^t$
be the smallest power of 2 greater than $p-1$, so $t$ is the
smallest integer greater than $\log_2 (p-1)$. Then FT modulo $2^t$
may be implemented in $O(t^2) = O((\log p)^2)$ steps. If we use FT
modulo $2^t$ in place of FT modulo $p-1$ in the above algorithm
then we will obtain a larger set of possible output pairs
$(l_1,l_2)$ with varying probabilities. However as in the case of
factoring, these pairs will lie with high probability sufficiently
near to the ``good'' pairs $(l_1, -yl_1)$ where $l_1$ is coprime
to $p-1$, so that $y$ may still be determined. The details of
dealing with the nearby pairs and assessing their probabilities,
are quite involved and given in \cite{SHO94}.

\section{The abelian hidden subgroup problem}

Given the above developments it is exciting to observe that the
concept of periodicity and the construction of the Fourier transform
may be generalized to apply to {\em any} finite group $G$. Our
discussion so far pertains simply to the special cases of the
additive group of integers modulo $N$ (for factoring) and the product
group $\cz_{p-1}\times \cz_{p-1}$ (for evaluating discrete
logarithms). The generalized viewpoint will also provide considerable
insight into the workings of the Fourier transform. We will now
outline the essential ideas involved restricting attention in this
section to the case of finite abelian groups.

Let $G$ be any finite abelian group. Let $f:G\rightarrow X$ be a
function on the group (taking values in some set $X$) and consider
\beq \label{KPER} K= \{ k\in G : f(k+g) = f(g) \mbox{ for all $g\in
G$} \} \eeq (Note that we write the group operation in additive
notation). $K$ is necessarily a subgroup of $G$ called the stabilizer
or symmetry group of $f$. It characterizes the periodicity of $f$
with respect to the group operation of $G$. For factoring where $G$
was ${\cal Z}_N$, $K$ was the cyclic subgroup of all multiples of
$r$.

The condition (\ref{KPER}) is equivalent to saying that $f$ is
constant on the cosets of $K$ in $G$. (Recall that the cosets are
subsets of $G$ of the form $g+K= \{ g+k:k\in K \}$ and they partition
all of $G$ into disjoint parts of equal size $|K|$).

Given a device that computes $f$, our aim is to suitably determine
the ``hidden subgroup'' $K$ e.g. we may ask for a set of
generators for $K$ or for an algorithm that outputs a randomly
chosen element of $K$. More precisely we wish to obtain this
information in time $O({\rm poly}(\log |G|))$ where $|G|$ is the
size of the group and the evaluation of $f$ on an input counts as
one computational step. (Note that we may easily determine $K$ in
time $O({\rm poly}( |G|))$ by simply evaluating and examining all
the values of $f$). We begin as in our examples by constructing
the state
\[ \ket{f}= \frac{1}{\sqrt{|G|}} \sum_{g\in G} \ket{g}\ket{f(g)}  \]
and read the second register. Assuming that $f$ is suitably
non-degenerate -- in the sense that $f(g_1 ) = f(g_2 )$ iff $g_1 -
g_{2} \in K$ i.e. that $f$ is one-to-one within each period -- we
will obtain in the first register
\begin{equation} \label{C3S2peri}
\ket{\psi (g_0 )} = \frac{1}{\sqrt{|K|}} \sum_{k\in K} \ket{g_0 +k}
\end{equation}
corresponding to seeing $f(g_0 )$ in the second register and $g_0$
has been chosen at random. In eq. (\ref{C3S2peri}) we have an equal
superposition of labels corresponding to a randomly chosen coset of
$K$ in $G$. Now $G$ is the disjoint union of all the cosets so that
if we read the label in eq. (\ref{C3S2peri}) we will see a random
element of a random coset, i.e. a label chosen equiprobably from all
of $G$, yielding no information at all about $K$.

The general construction of a ``Fourier transform on $G$'' will
provide a way of eliminating $g_0$ from the labels (just as in the
case of ${\cal Z}_N$) and the resulting state will then provide
direct information about $K$. Let $\cal H$ be a Hilbert space with a
basis $\{ \ket{g}: g\in G \} $ labeled by the elements of $G$. Each
group element $g_1 \in G$ gives rise to a unitary ``shifting''
operator $U(g_1 )$ on $\cal H$ defined by
\[ U(g_1 ) \ket{g}= \ket{g+g_1 } \hspace{1cm} \mbox{ for all $g$} \]
For any coset $g_0 +K$ let us write $\ket{g_0 +K}$ for the uniform
superposition $\frac{1}{\sqrt{|K|}} \sum_{k\in K} \ket{g_0 +k}$. Note
that the state in eq. (\ref{C3S2peri}) may be written as a
$g_0$-shifted state:
\begin{equation} \label{C3S219}
 \ket{g_0 +K} = U(g_0 ) \ket{K}
 \end{equation}

Our basic idea now is to introduce into $\cal H$ a new basis $\{
\ket{\chi_g }: g\in G \}$ of special states which are {\em
shift-invariant} in the sense that
\[ U(g_1 ) \ket{\chi_{g_2}} = e^{i\phi (g_1 , g_2 )} \ket{\chi_{g_2}}
\hspace{1cm} \mbox{ for all $g_1,g_2$ } \] i.e. the $\ket{\chi_g }$'s
are the common eigenstates of all the shifting operations $U(g)$.
Note that the $U(g)$'s all commute (since the group is abelian) so
such a basis of common eigenstates is guaranteed to exist. Then
according to eq. (\ref{C3S219}) if we view   $ \ket{K}$ and $\ket{g_0
+K}$ in the new basis, they will contain the same pattern of labels
determined by the subgroup $K$ only, and corresponding amplitudes
will differ only by phase factors. Thus the probability distribution
of the outcomes of a measurement in the new basis will directly
provide information about the subgroup $K$. More precisely it may be
shown \cite{EKE98} (and cf below) that this measurement provides a
uniform random sample from the so-called dual group of $K$ in $G$.

The Fourier transform $\cal F$ on $G$ is defined to simply be the
unitary transformation which rotates the shift invariant basis
back to the standard basis:
\[ {\cal F}\ket{\chi_g} = \ket{g} \hspace{1cm}\mbox{for all $g$} \]
Hence to read $\ket{\psi (g_0 )}$ in the new basis we just apply
$\cal F$ and read in the standard basis.

To give an explicit construction of $\cal F$ it suffices to give
the states $\ket{\chi_g}$ written as components in the standard
basis. There is a standard way of calculating these components
based on constructions from group representation theory. An
introduction with further references is given in
\cite{JOZ98b,EKE98} and here we will summarize the main points. If
we write \beq \label{IRREP} \ket{\chi_l} = \frac{1}{\sqrt{|G|}}
\sum_g \chi_l (g) \ket{g}\hspace{4mm} \mbox{ for each $l\in G$}
\eeq then we can take the functions $\chi_l : G \rightarrow {\cal
C}$ to be the $|G|$ irreducible representations of the group $G$.
Then the basic theorems of group representation theory (cf for
example \cite{JOZ98b}) guarantee that the states $\ket{\chi_l}$
are orthonormal and have the required shift invariant property.
Indeed shift invariance is a direct consequence of the basic
defining property of a representation: $\chi (g_1 + g_2 ) =
\chi(g_1 ) \chi (g_2 )$. For the group ${\cal Z}_N$ the
irreducible representations are given by $\chi_k (j) = \exp 2\pi
i\, jk/N$ for $j,k \in {\cal Z}_N$ and
\[ \ket{\chi_k}= \frac{1}{\sqrt{N}} \sum_{j=0}^{N-1}
e^{2\pi i\frac{jk}{N}} \ket{j} \] leading to the Fourier transform
formula given in eq. (\ref{C3S2FT}).

Which labels $l$ appear in ${\cal F}\ket{g_0 +K}$? It suffices to
consider ${\cal F}\ket{K}$ and from eq. (\ref{IRREP}) we get
directly
\[ {\cal F}\ket{K} = \frac{1}{\sqrt{|G|}\sqrt{|K|}} \sum_{l\in G}
\left( \sum_{k\in K} \chi_l(k) \right) \ket{l} \] Now, for Abelian
groups, the restriction of $\chi_l$ from $G$ to $K$ is an irreducible
representation of $K$ and the orthogonality relations for irreducible
representations give that $\sum_{k\in K} \chi_l (k) =0$ for all
$\chi_l$'s except the trivial representation defined by $\chi_l(k)
=1$ for all $k\in K$. In the latter case we have $\sum_{k\in K}
\chi_l(k) = |K|$. Hence ${\cal F}\ket{K}$ is a uniform superposition
of the $|G|/|K|$ labels $l$ such that $\chi_l$ restricts to the
trivial representation on $K$. If $K$ has a generator $r$ then the
latter condition is equivalent to $\chi_l(r) =1$ as we saw in the
example of factoring and discrete logarithms (where $r=(y,1)$). Thus
we are able to uniformly sample from this set of labels, which
distinguishes the possible $K$'s. This completes the quantum part of
the algorithm but to convert this into an explicit description of $K$
(say an actual set of generators) we need to use further mathematical
properties of $G$ e.g. properties of co-primality as illustrated in
our examples.

The above group-theoretic framework serves to generalize and extend
the applicability of the quantum algorithm for periodicity
determination. For example Simon considered the following problem:
suppose that we have a black box which computes a function $f$ from
$n$-bit strings to $n$-bit strings. It is also promised that the
function is ``two-to-one'' in the sense that there is a fixed $n$-bit
string $\xi$ such that
\begin{equation} \label{C3S2SI}
f(x+\xi )=f(x) \hspace{1cm} \mbox{for all $n$-bit strings $x$.}
\end{equation}
(Here $+$ denotes binary bitwise addition of $n$ bit strings.) Our
problem is to determine $\xi$.

To see that this is just a generalized periodicity determination
note that in the group $({\cal Z}_2 )^n$ of $n$-bit strings, every
element satisfies $ x+x=0 $. Hence eq. (\ref{C3S2SI}) states just
that $f$ is periodic on the group with periodicity subgroup $K= \{
0, \xi \}$. Thus to determine $\xi$ we construct the Fourier
transform on the group of $n$-bit strings and apply the standard
algorithm above. The relevant Hilbert space $\cal H$ with a basis
labeled by $n$-bit strings is just a row of $n$ qubits. The
irreducible representations of the group ${\cal Z}_2^N$ are the
functions $f_x (y) = (-1)^{x_1 y_1 }\ldots (-1)^{x_n y_n }$ where
$x = x_1 \ldots x_n$ and $y=y_1 \ldots y_n$ are n bit strings.
Thus the Fourier transform may be easily seen \cite{EKE98} to be
just the application of the 1-qubit Hadamard transform:
\[ H = \frac{1}{\sqrt{2}}
\left( \begin{array}{cc}  1& 1 \\ 1 & -1 \end{array} \right) \]
to each of the $n$ qubits.  The resulting quantum algorithm for
determining the hidden subgroup then reproduces Simon's original
algorithm \cite{SIM94}. It determines $\xi$ in $O(n^2 )$ steps
whereas it may be argued \cite{SIM94} that any classical algorithm
must evaluate $f$ at least $O(2^n )$ times.

\section{Non-abelian groups}

We will now consider the hidden subgroup problem in the situation
where $G$ and the subgroup $K$ may be non-abelian i.e. we have
$f:G\rightarrow X$ which is constant on the (left) cosets of $K$ in
$G$. We now also write the group operation multiplicatively. As
before our algorithm begins in the same way by producing the state
$\ket{g_0 K}$ where $g_0$ has been chosen at random. The passage from
abelian to non-abelian groups is accompanied by various potential
conceptual problems:

(a) (Construction of non-abelian Fourier transform). For abelian
groups the irreducible representations are always one dimensional
(i.e. the functions $\chi_l$ in eq. (\ref{IRREP})) whereas for
non-abelian groups they are functions $\chi : G\rightarrow U(d)$
taking values in the set $U(d)$ of all $d \times d$ unitary
matrices for suitable values of $d$. According to a basic theorem
of group representation theory \cite{FULHAR}, if $d_1 \ldots ,d_m$
are the dimensions of a complete set of irreducible unitary
representations $\chi_1 , \ldots ,\chi_m$ then $d_1^2 + \ldots
+d_m^2 = |G|$. Let us write $\chi_{i,jk} (g)$ for the $(j,k)$th
component of the unitary matrix $\chi_i (g)$. Thus as $i,j,k$ vary
we get $|G|$ complex valued functions and as in eq. (\ref{IRREP})
we may define the $|G|$ states:
\[ \ket{\chi_{i,jk}} = \frac{1}{\sqrt{|G|}} \sum_{g\in G} \chi_{i,jk}
(g) \ket{g}. \] The orthogonality relations of irreducible
representations \cite{FULHAR} guarantee that these are again
orthonormal states, called the Fourier basis, and the non-abelian
Fourier transform is defined as the unitary operation that rotates
this basis into standard position. In the abelian case, $j$ and
$k$ take only the value 1 and may be omitted. The Fourier basis
may be grouped into $m$ subsets of sizes $d_1^2 , \ldots ,d_m^2$
according to the value of $i$ and we may consider the associated
incomplete von Neumann measurement which distinguishes only the
various representations. We will denote this incomplete
measurement by ${\cal M}_{rep}$ and it will be important later (cf
(d) below).

(b) (Efficient implementation of non-abelian FT). For the
efficiency of our quantum algorithms it is important that FT be
implementable in poly$(\log |G| )$ computational steps. In the
abelian case this was a consequence of the FFT formalism.
Fortunately this formalism extends to the non-abelian case too
\cite{MAS95} requiring only that the group contains a suitable
tower of subgroups. For the standard FFT on $\cz_{2^n}$ this tower
is $H_0 \subset H_1  \subset \ldots \subset \cz_{2^n}$ where $H_k$
is the subgroup of multiples of $2^{n-k}$ in $\cz_{2^n}$. A
fundamental non-abelian group is the permutation group $G={\cal
P}_n$ on $n$ symbols. ${\cal P}_n$ contains the tower ${\cal P}_1
\subset {\cal P}_2 \subset \ldots \subset {\cal P}_n$ and its FT
has been shown to be efficiently implementable \cite{BEA97}.

(c) (Description of subgroups). Our quantum algorithm should
provide distinguishable outputs for different possible subgroups
$K$. In that case we say that the subgroup has been
information-theoretically determined. However in general it may
still be a difficult computational task to identify the actual
subgroup from the output result. For finite abelian groups a
fundamental structure theorem \cite{FRA} asserts that any such
group is isomorphic to a direct product of groups of the form
$\cz_n$. In this case any subgroup $K$ will have a simple
poly$(\log |G| )$ sized description given by a list of generators,
which we can require as the output of the algorithm. For
non-abelian groups the classification of possibilities is not so
simple. For example even the problem of deciding whether or not
two sets of generators and relations give isomorphic groups, is
known to be uncomputable! \cite{FRA}. Furthermore it is not
appropriate to ask for a list of all elements of $K$ as this may
be of size $O(|G|)$ i.e. exponentially large in $\log |G|$. We may
circumvent these difficulties of description by asking for less --
instead of characterising $K$ {\em per se}, we may for example ask
that the algorithm outputs a randomly chosen element of $K$ or
determines whether or not some chosen property of a subgroup holds
for the hidden subgroup.

(d) (Shift invariance). In the preceding section we used the
existence of the shift invariant basis $\ket{\chi_l}$ to give some
intuitive insight into why FT is useful for abelian hidden subgroups.
It provided a means of eliminating the effects of a randomly chosen
$g_0$ in the state $\ket{g_0 +K}$. The existence of a shift invariant
basis relies on the commuting of the shift operators $U(g)$ and this
is a consequence of the abelian-ness of $G$. In the non-abelian case
such a basis will not exist. However a restricted form of shift
invariance still survives because of the multiplicative property of
representations: $\chi_i (g_1 g_2 ) = \chi_i (g_1 ) \chi_i (g_2 ) $
(where the RHS is multiplication of $d_i \times d_i$ unitary
matrices). If we perform a complete measurement for the labels
$i,j,k$ (as in (a) above) on the state $\ket{gK}$ then the resulting
probability distribution will not be independent of $g$. However if
we perform the incomplete measurement ${\cal M}_{rep}$ then it is a
simple consequence \cite{HRT-S}  of the above multiplicative property
that the outcome distribution is independent of $g$, providing direct
(but generally incomplete) information about $K$ itself. (In the
abelian case this distribution is the uniform distribution over the
dual group of $K$ in $G$). In a similar way if $K$ and $L$ are {\em
conjugate} subgroups (i.e. $L=g_0 K g_0^{-1}$ for some $g_0$) then
any coset states $\ket{g_1 K}$ and $\ket{g_2 L}$ will also give
identical output distributions and hence the measurement ${\cal
M}_{rep}$ cannot distinguish conjugate subgroups. (In the abelian
case this is not a problem since subgroups are conjugate if and only
if they are equal).

There is no known efficient quantum algorithm that will solve the
hidden subgroup problem in general but we have various significant
partial results.

Let $G$ be any finite group and assume that the FT on $G$ can be
efficiently computed. Under this assumption, Hallgren, Russell and
Ta-Shma \cite{HRT-S} have shown that the hidden subgroup problem
may be efficiently solved for any {\em normal} subgroup $K$ of
$G$. We proceed as usual by first constructing a randomly chosen
coset state $\ket{g_0 K}$ (as in section 4) and then performing
the measurement ${\cal M}_{rep}$ in (a) (by performing FT and
reading the representation labels $i$ only). It is shown in
\cite{HRT-S} that $K$ may be reconstructed with high probability
from $O(\log |G|)$ repetitions of this procedure i.e. the $O(\log
|G| )$ measurement outcomes determine $K$ information
theoretically.

For abelian groups $G$ (where all subgroups are normal) this would
solve the general abelian hidden subgroup problem, except that FT
cannot be exactly implemented efficiently for a general abelian $G$.
Recall that in the examples of factoring and discrete logarithms we
needed to replace the Fourier transform by a slightly larger one --
in a dimension that was a power of 2 -- to take advantage of the FFT
formalism. This approximation to the true FT on $G$ was sufficiently
close to still allow the determination of the abelian hidden
subgroup. Kitaev \cite{KIT95} has described similar efficient
approximations to the FT on any abelian group which should suffice
for our purposes. Also, in view of (c) above, we could ask that the
algorithm in the abelian case determines $K$ more explicitly -- by
outputting an actual set of generators as in the examples of
factoring and discrete logarithms. Again this should be possible but
the detailed description of an efficient quantum algorithm for the
general abelian hidden subgroup problem seems not to have been
described in the literature (although the essential ingredients
appear to be implicit in the work of Kitaev \cite{KIT95} and Shor's
treatment \cite{SHO94} of factoring and discrete logarithms).

Returning to the most general hidden subgroup problem, Ettinger,
Hoyer and Knill \cite{EHK} have shown that $N= O(\log |G|)$
preparations of random coset states $\ket{g_1 K}, \ldots ,\ket{g_N
K}$ always suffice to determine $K$ information theoretically i.e.
there exists a quantum observable on the state $\ket{g_1 K}\otimes
\ket{g_2 K}\otimes \ldots \otimes \ket{g_N K}$ which will distinguish
all possible $K$'s with high probability (for any random choices of
$g_1, \ldots ,g_N$). However it is not known how to {\em efficiently}
implement such an observable in general. For the special case of
normal $K$'s the result of Hallgren, Russell and Ta-Shma gives
precisely such an efficiently implementable observable.

To conclude we will describe an important open question which can
be formulated as a non-abelian hidden subgroup problem. This is
the so-called graph isomorphism problem.

An (undirected) graph $A$ with $n$ vertices labeled $1,2, \ldots ,n$
may be described by an $n$ by $n$ matrix $M_A$ with entries that are
either 0 or 1. The $ij^{\rm th}$ entry is 1 if and only if the graph
has an edge joining vertices $i$ and $j$ (and we assume that $A$
always has at most one edge joining two vertices). Let ${\cal P}_n$
denote the group of all permutations of $n$ symbols $1,2, \ldots ,n$.
Two graphs $A$ and $B$ are said to be isomorphic if $B$ can be made
identical to $A$ by a re-labeling of its vertices i.e. if there
exists a permutation $\Pi \in {\cal P}_n$ such that $M_A$ is obtained
by simultaneously permuting the rows and columns of $M_B$ by $\Pi$.
The symmetry group of any graph $A$ on $n$ vertices is the subgroup
of all permutations $\Pi$ which leave $M_A$ unchanged when $\Pi$ is
applied to the rows and columns simultaneously. The graph isomorphism
problem is the following: given two connected graphs $A$ and $B$,
each on $n$ vertices, determine whether they are isomorphic or not.
We wish to perform this efficiently i.e. in poly$(n)$ steps. There is
no known efficient classical solution.

To re-formulate this problem as a hidden subgroup problem, let $C$
be the graph which is the disjoint union of $A$ and $B$, having
$2n$ vertices labeled $1,2, \ldots ,n,n+1, \ldots ,2n$ where $1,2,
\ldots ,n$ label $A$ and $n+1, \ldots , 2n$ label $B$. The
symmetry group $K$ of $C$ is evidently a subgroup of ${\cal
P}_{2n}$ but we can say more: since $A$ and $B$ are connected and
$C$ is the disjoint union, any symmetry of $C$ must either
separately permute the sets of labels $L_A = \{ 1,2, \ldots ,n \}$
and $L_B = \{ n+1, \ldots ,2n \}$ or else swap the two sets
entirely. Thus if $H$ denotes the group ${\cal P}_n \times {\cal
P}_n$ and $\sigma$ is the permutation of $1,2, \ldots ,2n$ that
swaps the two sets $S_A$ and $S_B$ in their listed order, then $K$
will always be a subset of the group $G=H \cup \sigma H$. $H$ is
the subgroup of $G$ containing all permutations that map $S_A$ and
$S_B$ into themselves whereas $\sigma H$ is its one other coset,
of all permutations that swap the elements of $S_A$ and $S_B$ (in
some arbitrary order). Now we may easily verify the following
facts:
\\(i) if $A$ and $B$ are not isomorphic then $K$ lies entirely in
$H$, \\ (ii) if $A$ and $B$ are isomorphic then exactly half of
the members of $K$ are in $H$ and half are in $\sigma H$.

Given any element $\Pi \in G$ it is easy to check whether it lies in
$H$ or $\sigma H$ (e.g. we just compute $\Pi(1)$ and check whether it
is $\leq n$ or $\geq n+1$). Hence we will have efficiently solved the
graph isomorphism problem if we are able to randomly sample from the
elements of $K$. This is a weak form of the hidden subgroup problem
in which we are not asking for the full information of $K$ but merely
whether it overlaps $\sigma H$ by half of its elements or is disjoint
from $\sigma H$, knowing that one of these two must always holds. In
our standard algorithm the function $f$ used to generate the random
coset state $\ket{g_0 K}$ is the efficiently computable
$f:G\rightarrow X$ where $X$ is the set of all matrices of size $2n
\times 2n$ with 0,1 entries and $f(\Pi)$ is the matrix obtained by
permuting the rows and columns of $M_C$ by $\Pi$.

Unfortunately none of the known partial results about efficient
quantum algorithms for determining hidden subgroups seem to apply
to this formulation of the graph isomorphism problem and the
possibility of an efficient solution remains an open challenge.
However given the already demonstrated success and mathematical
elegance of the Fourier transform formalism we can be optimistic
that an efficient algorithm might be derived along these lines.


\begin{thebibliography}{xx}


\bibitem{HAR65} Hardy, G. H. and Wright, E. M. (1965) {\em An Introduction
to the Theory of Numbers} (4th edition, Clarendon, Oxford).

\bibitem{SCH90} Schroeder, M. R. (1990) {\em Number Theory in Science and
Communication} (2nd enlarged edition, Springer, New York).

\bibitem{FULHAR} Fulton, M. and Harris, J. (1991) {\em Representation
Theory} (Springer GTM 29, Springer Verlag N.Y.)

\bibitem{FRA} Fraleigh, J. (1994) {\em A First Course in Abstract
Algebra} (5th edition, Addison-Wesley).

\bibitem{HRT-S} Hallgren, S., Russell, A. and Ta-Shma, A. (2000) Normal
subgroup reconstruction and quantum computing using group
representations, {\em Proc. 32nd Annual ACM Symposium on the Theory
of Computing -- STOC} (ACM Press, New York), 627-635.

\bibitem{SIM94} Simon, D. (1994) On the power of quantum computation,
 {\em Proc. of 35th Annual Symposium on the
Foundations of Computer Science}, (IEEE Computer Society, Los
Alamitos), p. 116 (Extended Abstract). Full version of this paper
appears in {\em S. I. A. M. Journal on Computing} (1997) {\bf 26},
1474-1483.

\bibitem{SHO94} Shor, P. (1994) Polynomial time algorithms for prime
factorisation and discrete logarithms on a quantum computer, {\em
Proc. of 35th Annual Symposium on the Foundations of Computer
Science}, (IEEE Computer Society, Los Alamitos), p. 124 (Extended
Abstract). Full version of this paper appears in {\em S. I. A. M.
Journal on Computing} {\bf 26} (1997), 1484-1510 and is also
available at quant-ph/9508027.

\bibitem{EKE96} Ekert, A. and  Jozsa, R. (1996) Quantum computation
and Shor's factoring algorithm, {\em Rev. Mod. Phys.} {\bf 68}, 733.

\bibitem{KIT95} Kitaev, A. (1995)  Quantum Measurements and
the Abelian Stabiliser Problem, preprint available at
http://xxx.lanl.gov/abs/quant-ph/9511026.


\bibitem{JOZ98b} Jozsa, R. (1998) Quantum algorithms and the fourier
transform, {\em Proc. Roy. Soc. London Ser A}, {\bf 454}, 323-337.

\bibitem{EKE98} Ekert, A. and Jozsa, R. (1998) Quantum algorithms:
entanglement enhanced information processing, {\em Phil. Trans. Roy.
Soc. London Ser A}, {\bf 356}, 1769-1782.

\bibitem{MAS95} Maslen, D. K. and Rockmore, D. N. (1995)
``Generalised FFT's -- A Survey of Some Recent Results'', in {\em
Proc. DIMACS Workshop on Groups and Computation -- II}.


\bibitem{BEA97} Beals, R. (1997) Quantum computation of fourier
transforms over symmetric groups, {\em Proc. 29th Annual ACM
Symposium on the Theory of Computing -- STOC} (ACM Press, New York),
48-53.

\bibitem{EHK} Ettinger, M., Hoyer, P. and Knill, E. (1999) Hidden
subgroup states are almost orthogonal, preprint available at
http://xxx.lanl.gov/abs/quant-ph/9901034.


\end{thebibliography}
\end{document}